\documentclass[12pt]{article}
\usepackage{amsmath, amssymb}
\usepackage[pdftex%,pagebackref,colorlinks%,linkcolor=red,anchorcolor=blue,citecolor=blue,urlcolor=brown
]{hyperref}

\numberwithin{equation}{section}
\def\p{\partial}

\begin{document}

\begin{titlepage}
\renewcommand{\thefootnote}{\fnsymbol{footnote}}

\begin{center}

\begin{flushright} \end{flushright}
\vspace{2cm}

\textbf{\Large{Mesons and Nucleons in Soft-Wall AdS/QCD \\[0.5cm]
                with Constrained Infrared Background}}\vspace{2cm}

Sheng Liu \hspace{0.2cm} and \hspace{0.2cm} Peng Zhang \\[0.5cm]

\textsf{E-mail: disney@emails.bjut.edu.cn,\\ 
\hspace{1.2cm} pzhang@bjut.edu.cn}\\[0.5cm]

\emph{Institute of Theoretical Physics,  College of Applied Sciences,  \\
      Beijing University of Technology,  Beijing 100124,  P.R.China}

\end{center}\vspace{1.5cm}

\centerline{\textbf{Abstract}}\vspace{0.5cm}

The purpose of this paper is to further study the soft-wall AdS/QCD model with constrained IR background proposed in \cite{Z0}.
By including a quartic bulk scalar potential we study various meson and nucleon spectra.
This model naturally realizes the asymptotical linearity of these mass spectra simultaneously,
together with correctly pattern of explicit and dynamical chiral symmetry breaking.
The agreement between the theoretical calculations and the experimental data is good.

\end{titlepage}
\setcounter{footnote}{0}

\section{Introduction}

It has been argued long ago, by 't Hooft \cite{tH}, that the large $N_{c}$ limit of quantum chromodynamics (QCD),
with fixed $g^2N_{c}$, should be described by its holographic dual string theory. This idea has been explicitly
realized by the AdS/CFT correspondence \cite{M, GKP, W}. In the infrared region QCD becomes strongly coupled, and
the effective dynamical degrees of freedom, instead of quarks and gluons, are hadrons in the particle zoo, like
$\pi$, $\rho$, $N$, etc. Therefore QCD cannot help us very much for understanding the properties low energy strong
interactions. However the idea of large $N_{c}$ expansions and holography supply a totally new point of view for
these hard and important problems. According to its general rule, when the 't Hooft coupling $g^2N_{c}$ is
large, we can use the effective theory in the bulk to study the strongly coupled dynamics of QCD.

Actually this has been an active region of research for recent years. Mainly there are two complementary ways to follow.
One is the top-down method, see e.g. \cite{KMMW, SS} which starts from some brane configurations in string theory.
It has the advantage of theoretical completeness, but the resulting model has only partial resemblances with the real QCD.
The other one is bottom-up, usually called AdS/QCD, see e.g. \cite{dTB1, EKSS, DP1}. This method assumes the bulk theory
living in the AdS$_5$ spacetime or its some IR deformation. The model contains several bulk fields, each of which
corresponds to a QCD operator, people uses observed experimental data and/or some properties of QCD, e.g. chiral symmetry
breaking, linear confinement, etc., to constrain the possible forms of the model. It supplies necessary conditions for a
would-be holographic theory of QCD should have. In this paper we will follow this bottom-up approach.

The so-called hard-wall model, defined on a slice of AdS$_5$ with a sharp IR cut-off, is developed first. These type of models
can correctly realize the pattern of chiral symmetry breaking and low-lying hadron states. For instance, scalar and pseudoscalar
mesons were studied in \cite{DP2}, tensor mesons in \cite{KLS}, and $b_1/h_1$ mesons in \cite{DHR}. Even hybrid exotic mesons
can be realized \cite{KK}. In addition to the meson sector, baryons can also be realized in the hard-wall model, see \cite{HIY, AHPS, PW1, Z1}.
However the main difficulty of the hard-wall approach is the absence of linear confinement. To remedy this drawback, A soft-wall model
is construct in \cite{KKSS}, which include a background dilaton field with quadratic growth at the deep IR region. By WKB-type arguments,
it can be shown that the excited meson spectrum exhibits the Regge behavior $m_n^2 \propto n+J$. In \cite{GKK}, by introducing a
quartic potential term for the bulk scalar, explicit and spontaneous chiral symmetry breaking are also correctly incorporated in
soft-wall AdS/QCD models. The relation with light-front dynamics is also discussed, see e.g. \cite{BdT1, dTB2}. A huge amount of
works have been done, a partial list is \cite{CDJN}-\cite{VSGL}.

In addition to the meson sector, various baryons also exhibit the approximate Regge behavior. One possible explanation \cite{Wil}
of this fact is that the baryon is composed of a quark and a diquark connected by a flux-tube string. So its structure is actually
similar with meson. In the literature of AdS/QCD, there are relatively few works considering the baryon linear spectrum, see
e.g. \cite{FBF, FK, VS1, BdT2}. In \cite{Z2}, and subsequently \cite{Z3}, we develop a soft-wall AdS/QCD model which realizes
asymptotically linear spectra for both mesons and nucleons. We achieve this by a cubic potential term for the bulk scalar and a new
parametrization of the its VEV. We also calculate the coupling between pion and nucleons. The main drawback of this model is that
the slopes of meson mass-squares are different, which is inconsistent with the real data. Unfortunately it is not easy to improve this.
Actually we will argue in this paper that it is impossible to get parallel meson slopes, while keeping the linear nucleon spectrum,
by only varying the scalar VEV or choosing different forms of the potential.

In \cite{Z0} we argue that, by requiring to have correct Regge-type spectrum in both meson and nucleon sectors, the IR asymptotic behavior of
various background fields in the model can be fully determined. The way around the above no-go theorem is to
allow the mass of bulk fields being $z$-dependent.\footnote{This has been suggested previously in e.g. \cite{CCW, VS2} for different purposes.}
This is actually very natural when considering possible anomalous dimension of the QCD operators. For operators which are not conserved currents,
like the quark condensates and baryon operators, the full conformal dimension is not the classical value. The anomalous part is
in general scale-dependent due to the running coupling constant, which translates to the $z$-dependence of the mass term for the
corresponding bulk fields according to the well-known mass-dimension relation. Therefore we make this assumption for the bulk scalar
field and the bulk Dirac field. We only require these masses approaching to the value dual to the classical dimension at
the UV boundary, since the high energy fixed-point of QCD is a free theory.

In the present paper we will further develop the model proposed in \cite{Z0}.
To be more realistic we include a quartic potential for the bulk scalar as in \cite{GKK}.
The trajectories of various meson sectors are parallel with each other, so improve the main drawback of our previous model in \cite{Z2, Z3}.
The remaining parts of this paper are organized as follows. In section 2 we discuss the meson sector of our model. It includes
scalar, vector, axial-vector, and pseudoscalar mesons. We compare the predicted masses and the corresponding data.
In section 3 we discuss the spin-1/2 nucleons.
We show that they also have asymptotically linear spectrum. We summarize this paper in section 4.

\section{Meson sector}

In our soft-wall AdS/QCD models,
all fields are defined in a five-dimensional Anti-de Sitter space with the metric
\begin{eqnarray}\label{ds2}
ds^2 = G_{MN}\, dx^M dx^N = a^2(z)(\eta_{\mu\nu}\, dx^\mu dx^\nu - dz^2),  \quad 0 < z < \infty \,.
\end{eqnarray}
The bulk action for the meson sector is:
\begin{eqnarray}\label{sm1}
S_M = \int d^4x\,dz\,\sqrt{G}\,e^{-\Phi}\left\{-\frac{1}{4g_5^2}(\|F_L\|^2+\|F_R\|^2)+\|DX\|^2-m_X^2\|X\|^2-\lambda\|X\|^4 \right\}.
\end{eqnarray}
Here $g_5^2=12\pi^2/N_c =  4\pi^2$ as usual.
$F_L$ and $F_R$ are the field strengths of the gauge potentials $L$ and $R$ respectively.
The covariant derivative is defined to be $D_M X = \partial_M X - iL_M X+iXR_M$, with $X$ in the bifundamental representation
of $SU(2)_L\times SU(2)_R$.
$\|X\|^2$  is the norm of the matrix $X$, i.e. $\|X\|^2 = \mathrm{Tr}(X^\dagger X)$.

\subsection{Background fields}
First we introduce the bulk scalar $X$,  it is assumed to have a z-dependent  VEV as follows:
\begin{eqnarray}\label{bsl1}
\langle{X}\rangle=\, \frac{1}{2}\, v(z) \begin{pmatrix} 1 & 0 \\ 0 & 1 \end{pmatrix}.
\end{eqnarray}
Then from the bulk action (\ref{sm1}) we get the equation that combine function $v(z)$ and the background dilaton $\Phi(z)$ :
\begin{eqnarray}
\p_z(a^3 e^{-\Phi}\p_z v)-a^5 e^{-\Phi}(m_X^2 v+\frac{\lambda}{2}v^3)=0\, . \label{EOMX}
\end{eqnarray}
We can deduce the mass-square $m^2_X$ may be $z$-dependent due to possible unusual dimension of $\overline{q}_Lq_R$.
Then according to (\ref{EOMX}),  $m^2_X$ can be expressed as :
\begin{equation}\label{mx22}
m^2_X = \frac{v''+(-\Phi'+3a'/a)v'}{a^2v}-\frac{\lambda}{2} v^2.
\end{equation}
The UV limit is still simple to argue, for the warp factor we have
\begin{equation}\label{az3}
a(z)\sim\frac{L}{z}, \qquad z\rightarrow0 .
\end{equation}
And for the scalar VEV we have
\begin{equation}\label{vz3}
v(z)\sim Az+Bz^3, \qquad z\rightarrow0 .
\end{equation}
By the mass-dimension relation $m^2_X = \Delta(\Delta-4)$,  we have
\begin{equation}\label{m2x3}
m^2_X(z)\sim-3, \qquad z\rightarrow0 .
\end{equation}
These  are the behaviors at UV boundary,  now we continue to study the IR situation .
For the dilaton it must be \cite{KKSS}
\begin{equation}
\Phi(z)\sim O(z^2), \qquad z\rightarrow \infty .
\end{equation}
which guarantees the mesons have linear spectra.
In order to obtain the spectral linearity of nucleons,  the IR limit of the warp factor is \cite{Z0}
\begin{equation}\label{az4}
a(z)\sim O(z), \qquad z\rightarrow \infty .
\end{equation}
To have parallel mass-square lines between vector and axial-vector mesons
we get the IR behavior of the scalar VEV
\begin{equation}\label{vz4}
v(z)\sim O(z^{-1}), \qquad z\rightarrow \infty.
\end{equation}

Now we use simple parametrization to smoothly connect these asymptotes from UV to IR as follows.
\begin{eqnarray}
\Phi(z)&=&\kappa^2 z^2\,,\label{psi4} \\[0.2cm]
a(z)&=&\frac{1+\mu z^2}{z}\,, \label{az2} \\
v(z)&=&\frac{Az+Bz^3}{1+Cz^4}\,.\label{vz2}
\end{eqnarray}
The parameters are determined by fitting the experimental data of
the pseudoscalar, scalar, vector and axial-vector meson masses. We take their values as
\begin{eqnarray}\label{6pms}
&&A=3.2\,\mathrm{MeV}, \quad  B=(394.5\,\mathrm{MeV})^3, \quad  C=(786.5\,\mathrm{MeV})^4\, ; \nonumber \\
&&\hspace{0.6cm}\mu=1153.6\,\mathrm{MeV}, \quad  \kappa=413.1\,\mathrm{MeV}, \quad  \lambda=5.99 \,.
\end{eqnarray}
In the following sections we will use them to calculate mass spectra and compare with the experimental data.

\subsection{Quadratic order action}

To get the fluctuation filed we define
\begin{eqnarray}
X\,=\,\left(\frac{v}{2}+S\right)e^{2iP},
\end{eqnarray}
here $S$ is a real scalar and \(P\) is a real pseudoscalar, and $X, S, P$ are all $2\times2$ matrices.
Next we define
\begin{eqnarray}\label{vm_am}
V_M = \frac{1}{2}\left(L_M+R_M\right),\quad A_M = \frac{1}{2}\left(L_M-R_M\right).
\end{eqnarray}
We expand the action (\ref{sm1}) to the quadratic order of these new fields
\begin{eqnarray}\label{sm3}
S_M^{(2)} = \int d^4xdz\left(\mathcal{L}_{P, A_5}^{(2)} + \mathcal{L}_{S}^{(2)} + \mathcal{L}_{V}^{(2)} + \mathcal{L}_{A}^{(2)}\right)
\end{eqnarray}
Each of them is as follows
\begin{eqnarray}
\hspace{-1.5cm}&&\mathcal{L}^{(2)}_{P,A_5} = -\frac{1}{2}\,a^3v^2e^{-\Phi}P^a\,\p^2P^a-
               \frac{1}{2}\,a^3v^2e^{-\Phi}(\p_5P^a-A_5^a)^2-\frac{1}{2g_5^2}\,ae^{-\Phi}A_5^a\,\p^{2}A_5^a\,,\label{LPA}\\[0.2cm]
\hspace{-1.5cm}&&\mathcal{L}^{(2)}_S = -\frac{1}{2}\,a^3e^{-\Phi} S^a\left\{\p^{2}-
               \frac{1}{\,a^3e^{-\Phi}}\,\p_{5}(a^3e^{-\Phi}\p_5)+m_X^2a^2-\frac{3}{4}\lambda\,a^2v\right\}S^a\,,\label{LS}\\[0.2cm]
\hspace{-1.5cm}&&\mathcal{L}^{(2)}_{V} = -\frac{1}{2g_5^2}\,ae^{-\Phi}V_\mu^a\left\{-\eta^{\mu\nu}\p^2+\p^\mu\p^\nu+
               \frac{1}{ae^{-\Phi}}\,\p_5(ae^{-\Phi}\p_5)\eta^{\mu\nu}\right\}V_\nu^a\,,\label{LV}\\[0.2cm]
               %+\frac{1}{g_5^2}\,ae^{-\Phi}\p_5V_{\mu}^a\p^{\mu}V_5^a\,,
\hspace{-1.5cm}&&\mathcal{L}^{(2)}_{A} = -\frac{1}{2g_5^2}\,ae^{-\Phi}A_\mu^a\left\{-\eta^{\mu\nu}\p^2+\p^\mu\p^\nu+
               \frac{1}{ae^{-\Phi}}\,\p_5(ae^{-\Phi}\p_5)\eta^{\mu\nu}-g_5^2a^2v^2\eta^{\mu\nu}\right\}A_\nu^a\,.\label{LA}
               %+\frac{1}{g_5^2}\,ae^{-\Phi}\p_5A_{\mu}^a\p^{\mu}A_5^a\,,
\end{eqnarray}
Some cross-terms have been canceled by gauge fixing terms \begin{eqnarray}
\mathcal{L}_{\mathrm{G.F.}}&=&-\,\frac{ae^{-\Phi}}{\, 2g_5^2\, \xi_V}\left\{\p^{\mu}V_\mu^a-\frac{\xi_V}{\, ae^{-\Phi}}\, \p_5(ae^{-\Phi}V_5^a)\right\}^2\nonumber\\
&&-\,\frac{ae^{-\Phi}}{\, 2g_5^2\, \xi_A}\left\{\p^{\mu}A_\mu^a-\frac{\xi_A}{\, ae^{-\Phi}}\, \p_5(ae^{-\Phi}A_5^a)+g_5^2\, \xi_Aa^2v^2P^a\right\}^2\, .
\end{eqnarray}
By using the unitary gauge $\xi\rightarrow\infty$ as in \cite{DP1}, we have
\begin{eqnarray}
\p_5(ae^{-\Phi}V_5^a)&=&0\, , \\[0.2cm]
\p_5(ae^{-\Phi}A_5^a)&=&g_5^2\, a^3v^2e^{-\Phi}P^a\, .\label{AP}
\end{eqnarray}
we can write $P^a$ in terms of $A_5^a$. Then equation (\ref{LPA}) becomes
\begin{eqnarray}
\mathcal{L}^{(2)}_{A_5}=-\frac{1}{2g_5^2}\, ae^{-\Phi}A_5^a\, \p^2D^2A_5^a
     -\frac{1}{2}\, a^3v^2e^{-\Phi}(D^2A_5^a)(D^2A_5^a)\, , \label{LA5}
\end{eqnarray}
and the quadratic order differential operator $D^2$ is defined by
\begin{eqnarray}
D^2f=-\, \p_5\left(\frac{\p_5(ae^{-\Phi}f)}{\, g_5^2\, a^3v^2e^{-\Phi}}\right)+f\, .
\end{eqnarray}

\subsection{Scalar mesons}

Next we should use Kaluza-Klein expansion to get the 4D effective action:
\begin{eqnarray}
S(x, z)=\sum_{n=0}^{\infty}\, \phi^{(n)}(x)\, f_S^{(n)}(z)\, , \label{KKf}
\end{eqnarray}
where $f_S^{(n)}$'s are eigenfunctions of the following problem
\begin{eqnarray}\label{sturms1}
-\frac{1}{\, a^3e^{-\Phi}}\, \p_{5}(a^3e^{-\Phi}\p_5f_S^{(n)})+(\, m_X^2a^2-\frac{3}{4}\lambda\, a^2v)\, f_S^{(n)}=M_S^{(n)2}f_S^{(n)}\, .\label{Mf}
\end{eqnarray}
with the boundary conditions
\begin{center}
$f_S^{(n)}|_{z\rightarrow0}=0\, , \quad\quad f_S^{(n)}|_{z\rightarrow\infty}=0\,  .$
\end{center}
And the orthonormality condition is
\begin{eqnarray}
\int_0^\infty a^3e^{-\Phi}f_S^{(n)}f_S^{(n')}dz=\, \delta_{nn'}\, .
\end{eqnarray}
Then we insert (\ref{KKf}) into (\ref{LS}) and do the integration over the $z$-coordinate,
we get exactly an effective 4D action for a cluster of scalar fields $\phi^{(n)}$.
We can also transform the Sturm-Liouville equation (\ref{Mf}) into a Schr\"{o}dinger form as $-\psi_S^{(n)\prime\prime}+V_S\psi_S^{(n)}=M_S^{(n)2}\psi_S^{(n)}$ in which we introduce the $V_S$ below. By setting $f_S^{(n)}=e^{\omega_{_S}/2}\psi_S^{(n)}$  with $\omega_S=\Phi-3\log{a}$,
the effective potential $V_S$ for scalar mesons is
\begin{eqnarray}\label{vs2}
V_S=\frac{1}{4}\omega_S'^{\, 2}-\frac{1}{2}\omega_S''+m_X^2a^2+\frac{3}{4}\lambda\, a^2v^2\, .
\end{eqnarray}
Here we have
\begin{equation}\label{vs1}
V_S\sim O(z^2),  \qquad z\rightarrow\infty .
\end{equation}
due to the background dilaton.
The eigenvalue problem (\ref{sturms1}) cannot be solved analytically. We have to rely on numerical calculations.
We use the former parameters listed in (\ref{6pms})to calculate the scalar meson masses.
The result and comparison with experimental data are shown in Table \ref{massf}.
The agreement between the theoretical and experimental values is pretty well.
%Actually it is better than the model in \cite{Z0} which does not contain the bulk scalar potential.
%This is the main motivation to include a quartic potential in the bulk action.
%
%f:scalar mesons
%
\begin{table}[h]
\centering
\begin{tabular}{|c|c|c|c|c|c|c|c|c|}
\hline
$n$                &  0   &   1  &  2   &  3   &  4   &  5   &   6   &   7 \\
\hline
$m_{\mathrm{exp}}$ & 550 & 980 & 1350 & 1505 & 1724 & 1992 & 2103 &  2189 \\
\hline
$m_{\mathrm{th}}$  & 550 & 999 & 1301 & 1544 & 1753 & 1939 & 2108 &  2265 \\
\hline
error              &0.0\% &2.0\% &3.6\% &2.6\% &1.7\% &2.7\% &0.3\% &3.5\% \\
\hline
\end{tabular}
\caption{\small{The experimental and theoretical values of scalar meson masses.
The average error is 2.03\%.}}\label{massf}
\end{table}

\subsection{Pseudoscalar mesons}

Similarly,  we expand the field $A_5$ in terms of its KK modes
\begin{eqnarray}
A_5(x, z)=\sum_{n=0}^{\infty}\, \pi^{(n)}(x)\, f_P^{(n)}(z)\, , \label{KKp}
\end{eqnarray}
with $f_P^{(n)}$ being the eigenfunction of the differential operator $D^2$
\begin{eqnarray}
-\, \p_5\left(\frac{\p_5(ae^{-\Phi}f_P^{(n)})}{\, g_5^2\, a^3v^2e^{-\Phi}}\right)+f_P^{(n)}=
\frac{M_P^{(n)2}}{\, g_5^2a^2v^2}\, f_P^{(n)}\, , \label{Mp}
\end{eqnarray}
and with the boundary condition \cite{DP2}
\begin{eqnarray}
\p_5(ae^{-\Phi}f_P^{(n)})|_{z\rightarrow0}=0\, , \quad\quad
f_P^{(n)}|_{z\rightarrow\infty}=0\, .\label{bcp}
\end{eqnarray}
According to general theories of the Sturm-Liouville problem,   we can normalize $f_P^{(n)}$
by the following orthonormality relation
\begin{eqnarray}
\int_0^\infty \frac{e^{-\Phi}}{\, av^2}\, \, f_P^{(n)}f_P^{(n')}\, dz=\frac{g_5^4}{M_P^{(n)2}}\, \delta_{nn'}\, .
\end{eqnarray}
We can rewrite this eigenvalue problem in a Schr\"{o}dinger form as in the scalar meson field. Define
\begin{eqnarray}
p=\frac{1}{\, g_5^2\, a^3v^2e^{-\Phi}}\, , \quad\quad
q=\frac{1}{\, ae^{-\Phi}}\, ,
\end{eqnarray}
and $\psi_P^{(n)}=ae^{-\Phi}{p}^{1/2}f_P^{(n)}$,  which satisfies the Schr\"{o}dinger equation
$-\psi_P^{(n)\prime\prime}+V_P\psi_P^{(n)}=M_P^{(n)2}\psi_P^{(n)}$ with the effective potential
\begin{eqnarray}\label{vp2}
V_P=\frac{\, 2pp''-p'^{2}+4pq}{4p^2}
\end{eqnarray}
Here we also have :
$$V_P\sim O(z^2),  \qquad z\rightarrow\infty$$
Then we find the asymptotical spectrum is linear with respect to the radial quantum number $n$.
The resulting mass spectra are listed in Table \ref{massp}.
%p:pseudoscalar mesons
%
\begin{table}[h]
\centering
\begin{tabular}{|c|c|c|c|c|c|}
\hline
$n$                &  0   &   1  &  2   &  3   &  4  \\
\hline
$m_{\mathrm{exp}}$ & 139 & 1300 & 1816 & 2070 & 2360 \\
\hline
$m_{\mathrm{th}}$  & 139 & 1662 & 1860 & 2040 & 2204 \\
\hline
error              &0.0\% &27.9\% &2.4\% &1.5\% &6.6\% \\
\hline
\end{tabular}
\caption{\small{The experimental and theoretical values of pseudoscalar meson masses.
The average error is 7.67\%.}}\label{massp}
\end{table}

\subsection{Vector mesons}

With the same procedure as the scalar and pseudoscalar mesons,  the field $V_\mu$ is expanded as:
\begin{eqnarray}
V_\mu(x, z)=\sum_{n=0}^{\infty}\, \rho_\mu^{(n)}(x)\, f_V^{(n)}(z)\, , \label{KKr}
\end{eqnarray}
with $f_V^{(n)}$ being eigenfunctions of the following problem
\begin{eqnarray}
&&-\frac{1}{\, ae^{-\Phi}}\, \p_5(ae^{-\Phi}\p_5f_V^{(n)})=M_V^{(n)2}f_V^{(n)}\, , \nonumber\\[0.2cm]
&&\hspace{0.3cm} f_V^{(n)}|_{z\rightarrow0}=0\, , \quad\quad f_V^{(n)}|_{z\rightarrow\infty}=0\, .\label{Mr}
\end{eqnarray}
We normalize $f_V^{(n)}$ by the following orthonormality condition
\begin{eqnarray}
\int_0^\infty ae^{-\Phi}f_V^{(n)}f_V^{(n')}dz=\, \delta_{nn'}\, .
\end{eqnarray}

Then we can get the effective 4D action for a tower of massive vector fields $\rho_\mu^{(n)}$,
which can be identified as the fields of $\rho$ mesons, with inserting (\ref{KKr}) into (\ref{LV}) and integrating over the $z$-coordinate.
Then we also transform (\ref{Mr}) into a Schr\"{o}dinger form,  by setting
$f_V^{(n)}=e^{\omega/2}\psi_V^{(n)}$ with $\omega=\Phi-\log{a}$.
The effective potential $V_V$ for vector mesons is
\begin{eqnarray}\label{vv2}
V_V=\frac{1}{4}\omega'^{\, 2}-\frac{1}{2}\omega''\, .
\end{eqnarray}
It is also of order $O(z^2)$ in the deep IR region,  i.e.($z\rightarrow \infty $),  and gives us asymptotically linear spectra for vector mesons.
The resulting mass spectra are listed in Table \ref{massr}.
%
%r:vector mesons
%
\begin{table}[h]
\centering
\begin{tabular}{|c|c|c|c|c|c|c|c|}
\hline
$n$                &  0   &   1  &  2   &  3   &  4   &  5   &   6  \\
\hline
$m_{\mathrm{exp}}$ & 775.5 & 1465 & 1570 & 1720 & 1909 & 2149 & 2265 \\
\hline
$m_{\mathrm{th}}$  & 982.9 & 1288 & 1533 & 1743 & 1930 & 2100 & 2257 \\
\hline
error              &26.8\% &12.1\% &2.4\% &1.3\% &1.1\% &2.3\% &0.4\% \\
\hline
\end{tabular}
\caption{\small{The experimental and theoretical values of vector meson masses.
The average error is 6.61\%.}}\label{massr}
\end{table}

\subsection{Axial-vector mesons}

We expand the field $A_\mu$ in terms of its KK modes
\begin{eqnarray}
A_\mu(x, z)=\sum_{n=0}^{\infty}\, a_\mu^{(n)}(x)\, f_A^{(n)}(z)\, , \label{KKa}
\end{eqnarray}
with $f_A^{(n)}$ being eigenfunctions of the following problem
\begin{eqnarray}
&&-\frac{1}{\, ae^{-\Phi}}\, \p_5(ae^{-\Phi}\p_5f_A^{(n)})+g_5^2\, a^2v^2f_A^{(n)}=M_A^{(n)2}f_A^{(n)}\, , \nonumber\\[0.2cm]
&&\hspace{0.8cm} f_A^{(n)}|_{z\rightarrow0}=0\, , \quad\quad f_A^{(n)}|_{z\rightarrow\infty}=0\, .\label{Ma}
\end{eqnarray}
The orthonormality condition for $f_A^{(n)}$ is :
\begin{eqnarray}
\int_0^\infty ae^{-\Phi}f_A^{(n)}f_A^{(n')}dz=\, \delta_{nn'}\, .
\end{eqnarray}
just as the same as vector mesons.
Again, insert (\ref{KKa}) into (\ref{LA}) and do the integration over the $z$-coordinate,
we get exactly an effective 4D action for a tower of massive axial-vector fields $a_\mu^{(n)}$,
As what we do in the former steps, we rewrite (\ref{Ma}) in a Schr\"{o}dinger form,  by setting
$f_A^{(n)}=e^{\omega/2}\psi_A^{(n)}$ with $\omega=\Phi-\log{a}$.
The effective potential $V_A$ for axial-vector mesons is
\begin{eqnarray}
V_A=\frac{1}{4}\omega'^{\, 2}-\frac{1}{2}\omega''+g_5^2\, a^2v^2\, . \label{VA}
\end{eqnarray}
It also has the quadratic behavior at $z\rightarrow\infty$,  and asymptotically linear spectra follows.
Note also that the last term is only $O(1)$, so the first term dominate which results in the same spectral slope
with that vector mesons. Having  different slopes is a main drawback of our previous model \cite{Z2, Z3}.
Now it has been removed by the proper choice of background fields.
The theoretical and the experimental values of axial-vector mesons
are listed in Table \ref{massa}.
%
%a:axial-vector mesons
%
\begin{table}[h]
\centering
\begin{tabular}{|c|c|c|c|c|c|c|}
\hline
$n$                &  0   &   1  &  2   &  3   &  4   &  5  \\
\hline
$m_{\mathrm{exp}}$ & 1230 & 1647 & 1930 & 2096 & 2270 & 2340  \\
\hline
$m_{\mathrm{th}}$  & 1438 & 1647 & 1844 & 2023 & 2188 & 2340  \\
\hline
error              &16.9\% &0.0\% &4.4\% &3.5\% &3.6\% &0.0\% \\
\hline
\end{tabular}
\caption{\small{The experimental and theoretical values of axial-vector meson masses.
The average error is 4.75\%.}}\label{massa}
\end{table}

\section{Nucleon sector}

To realize the spin-1/2 nucleon in the AdS/QCD, we can introduce two 5D Dirac spinors $\Psi_{1, 2}$ in the bulk as suggested in \cite{HIY}.
Each of them is also a isospin doublet. They are charged under the gauge fields $L_M$ and $R_M$ respectively. The action of nucleon sector is
\begin{eqnarray}
S_N&=&\int d^5x\,  \sqrt{G}\, \left(\mathcal{L}_K+\mathcal{L}_I\right) \, ,  \nonumber\\[0.1cm]
\mathcal{L}_K&=&i\overline{\Psi}_1\Gamma^M\nabla_M\Psi_1+i\overline{\Psi}_2\Gamma^M\nabla_M\Psi_2
                -m_{\Psi}\overline{\Psi}_1\Psi_1+m_{\Psi}\overline{\Psi}_2\Psi_2 \, ,  \\[0.2cm]
\mathcal{L}\, _I&=&-g\, \overline{\Psi}_1X\Psi_2-g\, \overline{\Psi}_2X^\dag\Psi_1\, .\label{SN} \nonumber
\end{eqnarray}
Here we have $\Gamma^M=e^M_A\Gamma^A=z\delta^M_A\Gamma^A$ with $\{\Gamma^A, \Gamma^B\}=2\eta^{AB}$.
We choose $\Gamma^A=(\gamma^a,  -i\gamma^5)$ with $\gamma^5=\mathrm{diag}(I, -I)$.
The covariant derivatives for spinors are
\begin{eqnarray}
\nabla_M \Psi_1&=&\p_M\Psi_1+\frac{1}{2}\, \omega^{AB}_M\Sigma_{AB}\Psi_1-iL_M\Psi_1 \, , \label{psi5}\\
\nabla_M \Psi_2&=&\p_M\Psi_2+\frac{1}{2}\, \omega^{AB}_M\Sigma_{AB}\Psi_2-iR_M\Psi_2 \, .\label{psi6}
\end{eqnarray}
Here $\Sigma_{AB}=\frac{1}{4}[\Gamma_A, \Gamma_B]$,  and the nonzero components of the spin
connection $\omega^{AB}_M$ is $\omega^{a5}_\mu=-\omega^{5a}_\mu=\frac{1}{z}\, \delta^a_\mu$.

\subsection{Nucleon spectrum}
The second order action is
\begin{eqnarray}
S_N^{(2)}&=&\int d^5x\,  \sqrt{G}\, \, (\, \mathcal{L}_K^{(2)}+\mathcal{L}_I^{(2)}) \, ,  \nonumber\\[0.1cm]
\mathcal{L}_K^{(2)}&=&\frac{1}{a}\, \sum_{i=1, 2}\,  \overline{\Psi}_i\left(i\gamma^\mu\p_\mu+
  \gamma^5\p_5+ \frac{2a'}{a}\gamma^5-m_{\Psi}a\right)\Psi_i\, , \\[0.1cm]
\mathcal{L}_I^{(2)}&=&\, -\frac{1}{2}\, gv\left(\, \overline{\Psi}_1\Psi_2+\overline{\Psi}_2\Psi_1\right)\, .\label{SN2} \nonumber
\end{eqnarray}
and we expand $\Psi_{1, 2}$ in terms of their KK modes :
\begin{eqnarray}
\Psi_1(x, z)=\begin{pmatrix} \sum_n N_{L}^{(n)}(x)\, f_{1L}^{(n)}(z)\, \,  \\[0.2cm] \sum_n N_{R}^{(n)}(x)\, f_{1R}^{(n)}(z)\, \, \end{pmatrix}\, \, , \quad
\Psi_2(x, z)=\begin{pmatrix} \sum_n N_{L}^{(n)}(x)\, f_{2L}^{(n)}(z)\, \,  \\[0.2cm] \sum_n N_{R}^{(n)}(x)\, f_{2R}^{(n)}(z)\, \, \end{pmatrix}\, \, .\label{PsiKK}
\end{eqnarray}
Here the $N^{(n)}_{L, R}$ are two-component objects,  which will be interpreted as the left-handed and right-handed parts
of a tower of 4D nucleon fields respectively,  that means £º
\begin{equation}
N^{(n)}(x)=(N^{(n)}_L, N^{(n)}_R)^{\mathrm{T}}
\end{equation}
when reducing to a 4D effective action.
And we have the following equations which the four internal functions $f^{(n)}$ satisfy
\begin{eqnarray}
%%%%%%%%%%%%%%%%%%%%%%%%%%%%%%%%%%%%%%%%%%%%%%%%%%%%%%%%%%%%%%%%%%%%%%%%%%%%%%%%%%%%%%%%%%%%%%%%%%%%%%%
\begin{pmatrix} \p_z-m_\Psi a+2a'/a & -u(z) \\[0.2cm] -u(z) & \p_z+m_\Psi a+2a'/a \,\,\end{pmatrix}
\begin{pmatrix} f_{1L}^{(n)} \\[0.2cm] f_{2L}^{(n)} \end{pmatrix}&=&
-\,M_N^{(n)}\begin{pmatrix} f_{1R}^{(n)} \\[0.2cm] f_{2R}^{(n)} \end{pmatrix}\,\,, \label{EOM1}\\[0.3cm]
%%%%%%%%%%%%%%%%%%%%%%%%%%%%%%%%%%%%%%%%%%%%%%%%%%%%%%%%%%%%%%%%%%%%%%%%%%%%%%%%%%%%%%%%%%%%%%%%%%%%%%%
\begin{pmatrix} \p_z+m_\Psi a+2a'/a & u(z) \\[0.2cm] u(z) & \p_z-m_\Psi a+2a'/a \,\,\end{pmatrix}
\begin{pmatrix} f_{1R}^{(n)} \\[0.2cm] f_{2R}^{(n)} \end{pmatrix}&=&
+\,M_N^{(n)}\begin{pmatrix} f_{1L}^{(n)} \\[0.2cm] f_{2L}^{(n)} \end{pmatrix}\,\,. \label{EOM2}
%%%%%%%%%%%%%%%%%%%%%%%%%%%%%%%%%%%%%%%%%%%%%%%%%%%%%%%%%%%%%%%%%%%%%%%%%%%%%%%%%%%%%%%%%%%%%%%%%%%%%%%
\end{eqnarray}
with $u(z)=\frac{1}{2}g_{\mathrm{Y}}a(z)v(z)$.
Note that these equations are general for any form of
various background fields, so it is the generalization of the corresponding equations in \cite{HIY}.
The UV boundary conditions are \cite{HIY}
\begin{eqnarray}
f_{1L}^{(n)}(z\rightarrow0)=0\,,\quad
f_{2R}^{(n)}(z\rightarrow0)=0\,. \label{bcf0}
\end{eqnarray}
The IR condition is as in \cite{Z2}, which is proper for soft-wall models
\begin{eqnarray}
f_{1R}^{(n)}(z\rightarrow\infty)=0\,,\quad
f_{2L}^{(n)}(z\rightarrow\infty)=0\,.\label{bcfoo}
\end{eqnarray}
To reduce the 5D bulk action to 4D, we also need the following orthonormality condition
\begin{eqnarray}
\int_0^\infty a^4 f_{aL}^{(n)}f_{aL}^{(n')} dz=
\int_0^\infty a^4 f_{aR}^{(n)}f_{aR}^{(n')} dz=\,\delta_{nn'}\,.\label{ONn}
\end{eqnarray}
From (\ref{EOM1}) and (\ref{EOM2}) it can be seen that only two of $f$'s are linear independent
\begin{eqnarray}
f_{2L}^{(n)}=-\epsilon f_{1R}^{(n)}\, , \quad\quad
f_{2R}^{(n)}=\epsilon f_{1L}^{(n)}\, , \label{P}
\end{eqnarray}
where $\epsilon=\pm1$ is the 4D parity.
We can transform (\ref{EOM1}) and (\ref{EOM2}) into a two-component
vector-valued Sturm-Liouville problem for $f_L^{(n)}=(f_{1L}^{(n)},f_{2L}^{(n)})^{\mathrm{T}}$ or
$f_R^{(n)}=(f_{1R}^{(n)},f_{2R}^{(n)})^{\mathrm{T}}$.
We can further rewrite the vector-valued Sturm-Liouville problem for e.g. $f_{L}^{(n)}$
into a Schr\"{o}dinger form $-\chi_L^{(n)\prime\prime}+V_N\chi_L^{(n)}=M_N^{(n)2}\chi_L^{(n)}$
by setting $f_L^{(n)}=a^{-2}\chi_L^{(n)}$. The potential matrix $V_N$ is
\begin{eqnarray}
V_N=\begin{pmatrix}\,\, m_{\Psi}^2a^2+(m_{\Psi}a)'+u^2 & u' \\[0.2cm]
    u' & m_{\Psi}^2a^2-(m_{\Psi}a)'+u^2 \,\,\,\end{pmatrix}\,.\label{VN}
\end{eqnarray}
We also have similar equations for the right-handed fields.

Based on the similar arguments about the anomalous dimensions,
we parametrize the bulk spinor mass also as a function of $z$ as
\begin{eqnarray}\label{mphi4}
m_{\Psi} = \frac{\frac{5}{2}+\mu_1 z}{1+\mu_2 z} .
\end{eqnarray}
For we have the mass-dimensional relation for spinors
\begin{eqnarray}
m_{\Psi} = \Delta-2
\end{eqnarray}
So this parametrization gives the correct UV limit 5/2, corresponding to the classical dimension 9/2 of the baryon operator by the equation above.
And at IR $m_{\Phi}$ will tend to a constant $\mu_1 / \mu_2$,  and this is also reasonable.
By fitting the spin-1/2 nucleon mass we choose
\begin{eqnarray}\label{3pr1}
\mu_1 = 1.16 \,  \mathrm{GeV},  \quad \mu_2 = 7.8 \,  \mathrm{GeV},  \quad g_Y =8.74  .
\end{eqnarray}
The resulting mass spectra and the corresponding data are listed in Table \ref{massn}.
%
%Nucleon£º
%
\begin{table}[h]
\centering
\begin{tabular}{|c|c|c|c|c|c|c|c|}
\hline
$n$                &  0   &   1  &  2   &  3   &  4   &  5   &   6  \\
\hline
$m_{\mathrm{exp}}$ & 939 & 1440 & 1535 & 1650 & 1710 & 2090 & 2100 \\
\hline
$m_{\mathrm{th}}$  & 941 & 1402 & 1536 & 1767 & 1819 & 2026 & 2057 \\
\hline
error              &0.2\% &2.6\% &0.1\% &7.1\% &6.4\% &3.1\% &2.0\% \\
\hline
\end{tabular}
\caption{\small{The experimental and theoretical values of the spin-1/2 nucleon masses.
The average error is 3.06\%.}}\label{massn}
\end{table}

\section{Summary}

In this paper we further develop the model proposed in \cite{Z0}. The main motivation of this model is to
correctly reproduce the observed spectral pattern of both mesons and nucleons. In the original soft-wall model
the quadratic dilaton is introduced for the linear spectra of mesons. To further constrain the IR behavior of
other background fields, $a(z)$ and $v(z)$, we need to consider more spectral details. Two key facts which help
us to fix this is: (1) nucleons also have linear spectra, and (2) various meson sectors have the same spectral slopes.
Combining these two requires $a(z)\sim O(z)$ and $v(z)\sim O(z^{-1})$ as $z\rightarrow\infty$. In the present work
we include a quartic potential for the bulk scalar to improve our model, and carefully study the spectra of various
mesons and nucleons. The agreement between the theoretical calculation and the experimental data is rather good.
Actually it can be easily generalized to include more baryon sectors, e.g. the $\Delta$.
These discussions show a way to consistent to consider mesons and baryons simultaneously in one AdS/QCD model.
The problem that they need different IR cutoffs in the hard-wall model disappears here just by definition.
It is a proper setup to further study meson-baryon interactions in future works.

\section*{Acknowledgements}

We appreciate Prof. Y.-C. Huang for his encouragement.
SL would also like to thank  K. Zhao and Suzanna Meng for  insightful discussions,
and Prof. V. Ledoux for her help about the MATSLISE package.

\end{document}